\documentstyle[12pt,fleqn]{article}

\font\twlgot =eufm10 scaled \magstep1
\font\egtgot =eufm8
\font\sevgot =eufm7

\font\twlmsb =msbm10 scaled \magstep1
\font\egtmsb =msbm8
\font\sevmsb =msbm7

\newfam\gotfam
\def\pgot{\fam\gotfam\twlgot}
\textfont\gotfam\twlgot
\scriptfont\gotfam\egtgot
\scriptscriptfont\gotfam\sevgot
\def\got{\protect\pgot}

\newfam\msbfam
\textfont\msbfam\twlmsb
\scriptfont\msbfam\egtmsb
\scriptscriptfont\msbfam\sevmsb
\def\Bbb{\protect\pBbb}
\def\pBbb{\relax\ifmmode\expandafter\Bb\else\typeout{You cann't use
Bbb in text mode}\fi}
\def\Bb #1{{\fam\msbfam\relax#1}}

\let\Large=\large

\def\op#1{\mathop{{\it\fam0} #1}\limits}
\newcommand{\id}{{\rm Id\,}}

\newcommand{\Ker}{{\rm Ker\,}}
\newcommand{\im}{{\rm Im\, }}
\newcommand{\hm}{{\rm Hom\,}}
\newcommand{\dif}{{\rm Diff\,}}
\newcommand{\lng}{\langle}
\newcommand{\rng}{\rangle}

\newcommand{\eqref}[1]{(\ref{#1})}
\newcommand{\beq}{\begin{equation}}
\newcommand{\eeq}{\end{equation}}
\newcommand{\ben}{\begin{eqnarray}}
\newcommand{\een}{\end{eqnarray}}
\newcommand{\be}{\begin{eqnarray*}}
\newcommand{\ee}{\end{eqnarray*}}
\newcommand{\bea}{\begin{eqalph}}
\newcommand{\eea}{\end{eqalph}}

\newcommand{\cO}{{\got O}}
\newcommand{\cA}{{\cal A}}

\newcommand{\gd}{{\got d}}
\newcommand{\gj}{{\got J}}

\newcommand{\cH}{{\cal H}}

\newcommand{\cZ}{{\cal Z}}

\newcommand{\cK}{{\cal K}}

\newcommand{\cS}{{\cal S}}

\newcommand{\bL}{{\bf L}}
\newcommand{\bb}{{\bf 1}}

\newcommand{\bp}{{\bf p}}

\newcommand{\al}{\alpha}
\newcommand{\bt}{\beta}
\newcommand{\dl}{\delta}
\newcommand{\la}{\lambda}

\newcommand{\f}{\phi}

\newcommand{\om}{\omega}
\newcommand{\Om}{\Omega}
\newcommand{\m}{\mu}

\newcommand{\G}{\Gamma}

\newcommand{\ve}{\varepsilon}
\newcommand{\th}{\theta}

\newcommand{\si}{\sigma}

\newcommand{\w}{\wedge}

\newcommand{\wh}{\widehat}
\newcommand{\ol}{\overline}
\newcommand{\ul}{\underline}
\newcommand{\dr}{\partial}

\newcommand{\mar}[1]{}

\newcommand{\ar}{\op\longrightarrow}

\newcommand{\ot}{\otimes}

\newcounter{eqalph}
\newcounter{equationa}
\newcounter{example}
\newcounter{remark}
\newcounter{theorem}
\newcounter{proposition}
\newcounter{lemma}
\newcounter{corollary}
\newcounter{definition}

\def\theremark{\arabic{remark}}

\def\thedefinition{\arabic{definition}}

\newenvironment{proof}{\noindent {\bf Proof.}}{\hfill{\footnotesize\bf
QED} \bigskip }
\newenvironment{ex}{\refstepcounter{remark} \medskip\noindent{\bf Example
\theremark.}}{{\Large $\bullet$} \bigskip }
\newenvironment{rem}{\refstepcounter{remark} \medskip\noindent{\bf Remark
\theremark.}}{{\Large $\bullet$} \bigskip }
\newenvironment{theo}{\refstepcounter{definition} \bigskip\noindent{\sc
Theorem \thedefinition}.}{$\Box$\bigskip }
\newenvironment{prop}{\refstepcounter{definition} \bigskip\noindent{\sc
Proposition \thedefinition}.}{$\Box$ \bigskip }

\newenvironment{defi}{\refstepcounter{definition} \bigskip\noindent{\sc
Definition \thedefinition}.}{$\Box$ \bigskip }

\newenvironment{eqalph}{\stepcounter{equation}
\setcounter{equationa}{\value{equation}}
\setcounter{equation}{0}

\begin{eqnarray}}{\end{eqnarray}
\setcounter{equation}{\value{equationa}}}

\begin{document}

\begin{center}
{\large\bf JETS AND CONNECTIONS IN COMMUTATIVE AND NONCOMMUTATIVE GEOMETRY}
\bigskip

{\sc  LUIGI MANGIAROTTI}\footnote{E-mail: mangiaro@camserv.unicam.it}

{\small
 Department of Mathematics and Physics, University of Camerino, 

62032
Camerino (MC), Italy }
\medskip

{\sc GENNADI SARDANASHVILY}\footnote{E-mail: sard@grav.phys.msu.su}

{\small
 Department of Theoretical Physics, Physics Faculty, Moscow State
University,

 117234 Moscow, Russia}
\bigskip

\end{center}

\begin{small}

It is emphasized that equivalent definitions of connections on modules over a
commutative ring are not so in noncommutative geometry.
\end{small}
\bigskip

\section{Introduction}

The jet modules $\gj^k(P)$ of a module $P$ over a commutative ring $\cA$ are
well-known to be a representative object of linear differential operator on
$P$ \cite{vinb}. Furthermore, a connection on a module $\cA$ is defined to be
a splitting of the exact sequence 
\mar{+175}\beq
0\ar \cO^1\ot P\to \gj^1(P)\ar^{\pi^1_0} P\ar 0, \label{+175}
\eeq
where $\cO^1$ is the module of differentials of $\cA$. In the case of
structure modules of smooth vector bundles, these notions of jets and
connections coincide with those in differential geometry of fibre bundles
where connections on a fibre bundle $Y\to X$ are sections of the affine jet
bundle $J^1Y\to Y$ \cite{book}.
In general, the notion of jets of modules fails to be extended to modules over
a noncommutative ring $\cA$ since it implies a certain commutativity
property of a differential calculus $\cO^*$ over $\cA$. In relation to this
circumstance, we match different definitions of connections which being
equivalent for modules over a commutative ring are not so in noncommutative
geometry.

\section{Modules in noncommutative geometry}

Let $\cA$ be an associative unital algebra over a commutative ring
$\cK$, i.e., a $\cA$ is a $\cK$-ring. One
considers right [left]
$\cA$-modules and $\cA$-bimodules (or $\cA-\cA$-bimodules in the
terminology of \cite{mcl}). A bimodule $P$ over an algebra $\cA$ is
called a  central 
bimodule if 
\mar{+890}\beq
pa=ap, \qquad \forall p\in P, \qquad \forall a\in \cZ(\cA), \label{+890}
\eeq
where $\cZ(\cA)$ is the centre of the algebra $\cA$. By a  centre
 of a $\cA$-bimodile $P$  is called
a $\cK$-submodule $\cZ(P)$ of $P$ such that 
\be
pa\op=^{\rm def} ap, \qquad \forall p\in \cZ(P), \qquad \forall a\in \cA.
\ee
If $\cA$ is a commutative algebra,
every right [left] module $P$ over $\cA$ becomes canonically a central
bimodule by putting 
\be
pa=ap, \qquad \forall p\in P, \qquad \forall a\in \cA. 
\ee
If $\cA$ is a noncommutative
algebra, every right [left] $\cA$-module $P$ is also a
$\cZ(\cA)-\cA$-bimodule [$\cA-\cZ(\cA)$-bimodule] such that the equality
(\ref{+890}) takes place, i.e., it is a central $\cZ(\cA)$-bimodule.
From now on, by a
$\cZ(\cA)$-bimodule is meant a 
central $\cZ(\cA)$-bimodule. 
For the sake of
brevity, we say that, given an associative algebra $\cA$, 
right and left $\cA$-modules, central $\cA$-bimodules and
$\cZ(\cA)$-modules are 
$A$-modules of type $(1,0)$, $(0,1)$, 
$(1,1)$ and $(0,0)$, respectively, where $A_0=\cZ(\cA)$ and $A_1=\cA$.
Using this notation, let us recall a few basic operations with modules.
\begin{itemize}
\item If $P$ and $P'$ are $A$-modules of the same type $(i,j)$, so is its
direct sum $P\oplus P'$.
\item Let $P$ and $P'$ be $A$-modules of
types $(i,k)$ and $(k,j)$, respectively. Their tensor product $P\ot P'$
(see \cite{mcl}) defines an $A$-module of
type $(i,j)$.
\item Given an $A$-module $P$ of type
$(i,j)$, let $P^*=\hm_{A_i-A_j}(P,\cA)$ be its 
$\cA$-dual. One can show that $P^*$ is the module of type
$(i+1,j+1){\rm mod}\,2$ \cite{dub}. In particular, $P$ and
$P^{**}$ are $A$-modules of the same type.
There is the natural homomorphism
$P\to P^{**}$.
For instance, if $P$ is a projective module of finite rank, so is its
dual $P^*$ and  
$P\to P^{**}$ is an isomorphism \cite{mcl}.
\end{itemize}

There are several equivalent definitions of a projective module. One says
that a right [left] module $P$ is  projective if $P$ 
is a direct summand of a right [left] free module, i.e., there exists a
module $Q$ 
such that $P\oplus Q$ is a free module \cite{mcl}. Accordingly, a
module $P$ is projective if and only if $P=\bp S$ where $S$ is a free
module and $\bp$ is an idempotent, i.e., 
an endomorphism of $S$ such that $\bp^2=\bp$. 
We will refer to
projective $\Bbb C^\infty(X)$-modules of finite rank in connection with the
Serre--Swan theorem below. Recall that a module
is said to be of finite rank or simply  finite if it is a
quotient of a finitely generated free module.

Noncommutative geometry deals with unital complex involutive algebras
(i.e., unital $*$-algebras) as a rule. Let $\cA$ be such an algebra  (see
\cite{dixm}). It should be emphasized that one cannot use right or left
$\cA$-modules,
but only modules of type $(1,1)$ and $(0,0)$
since the involution of $\cA$ reverses the order of product in $\cA$. A central
$\cA$-bimodule $P$ over  $\cA$ is said to be a  $*$-module over 
a $*$-algebra $\cA$ if it is equipped with an antilinear involution
$p\mapsto p^*$ such that
\be
(apb)^*=b^*p^*a^*, \qquad \forall a,b\in\cA, \qquad p\in P.
\ee
A $*$-module is said to be a finite projective module if it is a finite
projective right [left] module.

As well-known, noncommutative geometry is developed
in main as a generalization of the calculus in commutative rings of
smooth functions.
Let $X$ be a locally compact topological space and $\cA$ a $*$-algebra
$\Bbb C^0_0(X)$ of complex continuous functions on $X$ which vanish at
infinity of $X$.  Provided with the norm
\be
||f||=\op{\rm sup}_{x\in X}|f|, \qquad f\in\cA,
\ee
this algebra is a $C^*$-algebra \cite{dixm}. Its
spectrum $\wh \cA$ is homeomorphic to $X$. Conversely, 
any commutative $C^*$-algebra
$\cA$ has a locally compact spectrum $\wh\cA$ and, in accordance with the
well-known  Gelfand--Na\u\i mark theorem, it is
isomorphic to the algebra $\Bbb C^0_0(\wh\cA)$ of complex continuous
functions on $\wh\cA$ which 
vanish at infinity of $\wh\cA$ \cite{dixm}. If $\cA$ is a unital
commutative $C^*$-algebra, its spectrum $\wh\cA$ is
compact. 
Let now $X$ be a compact manifold. The $*$-algebra $\Bbb
C^\infty(X)$ of smooth complex functions on $X$ is a dense 
subalgebra of the unital $C^*$-algebra $\Bbb C^0(X)$ of continuous functions
on $X$. This is
not a $C^*$-algebra, but it is a Fr\'echet algebra in its natural
locally convex topology of compact convergence for all derivatives. In
noncommutative geometry, one does not use the theory of locally convex
algebras (see \cite{mich}), but considers dense unital subalgebras of
$C^*$-algebras in a purely algebraic fashion. 

 Let $E\to X$ be a smooth $m$-dimensional complex
vector bundle over a 
compact manifold $X$. The module $E(X)$ of its global sections is a
$*$-module over the ring $\Bbb 
C^\infty(X)$ of smooth complex functions on $X$. It is a projective
module of finite rank. Indeed, let $(\f_1,\ldots,\f_q)$ be a smooth
partition of unity such that $E$ is trivial over the sets $U_\zeta\supset
{\rm supp}\,\f_\zeta$, together
with the transition functions $\rho_{\zeta\xi}$. Then
$p_{\zeta\xi}=\f_\zeta\rho_{\zeta\xi}\f_\xi$ are smooth $(m\times
m)$-matrix-valued functions on $X$. They satisfy 
\mar{+830}\beq
\op\sum_\kappa p_{\zeta\kappa}p_{\kappa\xi}=p_{\zeta\xi}, \label{+830}
\eeq
and so assemble into a $(mq\times mq)$-matrix $\bp$ whose entries are
smooth complex functions on $X$. Because of (\ref{+830}), we obtain
$\bp^2=\bp$. Then any section $s$ of $E\to X$ is represented by a column
$(\f_\zeta s^i)$ of smooth complex functions on $X$ such that $\bp s=s$. 
It follows that $s\in \bp\Bbb C(X)^{mq}$, i.e., $E(X)$ is a projective
module. The
above mentioned  Serre--Swan theorem
\cite{swan,var} provides a converse assertion.

\begin{theo} 
Let $P$ be
a finite projective $*$-module over $\Bbb C^\infty(X)$. There exists a complex
smooth vector bundle $E$ over 
$X$ such that $P$ is isomorphic to the module $E(X)$ of global sections
of $E$. 
\end{theo}

In noncommutative geometry,
one therefore thinks of a finite projective $*$-module over a dense
unital $*$-subalgebra of a $C^*$-algebra as being a  noncommutative
vector bundle. 

\section{Commutative differential calculus}

Let us summarize some basic facts on the differential
calculus in modules over a commutative $\cK$-ring $\cA$
\cite{vinb,book,kosz60}. 

Let $P$ and $Q$ be left $\cA$-modules. Right modules are studied in a
similar way. The set
$\hm_\cK(P,Q)$ of
$\cK$-module homomorphisms of $P$ into $Q$ is endowed with the
$\cA-\cA$-bimodule structure by the left and right multiplications
\mar{5.29}\beq
(a\f)(p) = a\f(p),  \qquad  (\f\star a)(p) = \f (a p),\qquad a\in
\cA, \qquad p\in P. \label{5.29}
\eeq
However, this is not a central $\cA$-bimodule because $a\f\neq \f\star a$ in
general.
 Let us denote 
\mar{+a10}\beq
\dl_a\f= a\f -\f\star a. \label{+a10}
\eeq
 
\begin{defi}\label{ch535} \mar{ch535} An element $\Delta\in\hm_\cK(P,Q)$ is
called an  $s$-order  linear differential operator from the
$\cA$-module $P$ to the $\cA$-module $Q$ if 
\be
\dl_{a_0}\circ\cdots\circ\dl_{a_s}\Delta=0 
\ee
for arbitrary collections of $s+1$ elements of $\cA$. It is also called a
$Q$-valued differential operator on $P$.  
\end{defi}

In particular, a first order linear differential operator
$\Delta$ obeys the condition
\mar{mos040}\beq
\dl_a\circ\dl_b\Delta)(p)=\Delta(abp)-a\Delta(bp) -b\Delta(ap)+
ab\Delta(p)=0
 \label{mos040}
\eeq
for all $p\in P$,  $b,c\in \cA$.

A  first order differential operator $\dr$ from
$\cA$ to an $\cA$-module $Q$ is called the $Q$-valued  derivation of the
algebra
$\cA$ if it obeys the 
 Leibniz rule
\mar{+a20}\beq
\dr(aa') = a\dr(a') + a'\dr(a), \qquad \forall a,a'\in \cA. \label{+a20}
\eeq
This is a particular condition (\ref{mos040}).

Turn now to the modules of jets.
Given an $\cA$-module $P$, let us consider the tensor product
$\cA\op\otimes_\cK P$ of $\cK$-modules provided with the left $\cA$-module
structure
\mar{5.43}\beq
b(a\otimes p)\op=^{\rm def}(ba)\otimes p, \qquad \forall b\in\cA. \label{5.43}
\eeq
For any $b\in \cA$, we introduce the left $\cA$-module morphism
\mar{+a5}\beq
\dl^b(a\otimes p)= (ba)\otimes p - a\otimes (b p). \label{+a5}
\eeq
Let  
$\m^{k+1}$ be the submodule of the left $\cA$-module $\cA\op\otimes_\cK P$ 
generated by all elements of the type 
\be
\dl^{b_0}\circ \cdots \circ\dl^{b_k}(\bb\otimes p). 
\ee

\begin{defi}\label{ch521} \mar{ch521}
The  $k$-order jet module of the
$\cA$-module $P$ 
is defined to be the quotient $\gj^k(P)$ of $\cA\otimes P$ by $\m^{k+1}$.
It is a left $\cA$-module with respect to the multiplication
\mar{+a21}\beq
b(a\ot p\,{\rm mod}\,\m^{k+1})=ba\ot p\,{\rm mod}\,\m^{k+1}. \label{+a21}
\eeq
\end{defi}

Besides the left $\cA$-module structure induced by
(\ref{5.43}), the
$k$-order jet module
$\gj^k(P)$ also admits the left $\cA$-module structure given by the
multiplication
\mar{mos057}\beq
b\star(a\otimes p\,{\rm mod}\,\m^{k+1})= a\otimes (bp)\,{\rm
mod}\,\m^{k+1}. \label{mos057}
\eeq
It is called the  $\star$-left module
structure.  There is the $\star$-left
$\cA$-module homomorphism 
\mar{5.44}\beq
J^k: P\to \gj^k(P), \qquad J^kp=\bb\otimes p\,{\rm mod}\,\m^{k+1},
\label{5.44}
\eeq
such that $\gj^k(P)$ as a left $\cA$-module is generated by
the elements $J^kp$,
$p\in P$. It is readily observed that
the homomorphism $\gj^k$ (\ref{5.44}) is a $k$-order differential operator
(compare the relation (\ref{mos040}) and the relation (\ref{mos041}) below). 

\begin{rem} \label{+a30} \mar{+a30}
If $P$ is a $\cA-\cA$-bimodule, the tensor product $\cA\op\ot_\cK P$ is also
provided with the right $\cA$-module structure
\be
(a\ot p)b\op=^{\rm def} a\ot pb, \qquad \forall b\in\cA,
\ee
and so is the jet module $\gj^k(P)$:
\be
(a\ot p\,{\rm mod}\,\m^{k+1})b=a\ot(pb) \,{\rm mod}\,\m^{k+1}. \label{+a31}
\ee
If $P$ is a central bimodule, i.e., 
\be
ap=pa, \quad \forall a\in \cA, \quad p\in P,
\ee
the $\star$-left $\cA$-module structure (\ref{mos057}) is equivalent to
the right $\cA$-module structure (\ref{+a31}).
\end{rem}

The jet modules possess the properties similar to those of jet manifolds.
In particular, since $\m^r\subset \m^s$, $r>s$, there is the
the inverse system of epimorphisms
\be
\gj^s(P)\ar^{\pi_{s-1}^s} \gj^{s-1}(P)\ar \cdots\ar^{\pi^1_0} P. 
\ee
Given the repeated jet module $\gj^s(\gj^k(P))$, there exists the
monomorphism $\gj^{s+k}(P)\to \gj^s(\gj^k(P))$. 

In particular, the first order jet module $\gj^1(P)$
consists of elements $ a\otimes p\,{\rm mod}\,\m^2$, i.e., elements
$a\ot p$ modulo the relations
\mar{mos041}\ben
&& \dl^a\circ \dl^b(\bb\ot p)=
\label{mos041}\\
&& \qquad (\dl_a\circ\dl_b\gj^1)(p)=\bb\ot(abp) -a\otimes (bp) -b\otimes
(ap) + ab\otimes p =0. \nonumber
\een
The morphism $\pi^1_0:\gj^1(P) \to P$ reads
\mar{+a13}\beq
\pi^1_0:a\ot p\,{\rm mod}\,\m^2\to ap. \label{+a13}
\eeq

\begin{theo}\label{ch526} \mar{ch526} For any differential operator $\Delta\in
\dif_s(P,Q)$ there is a unique homomorphism ${\got f}^\Delta: \gj^s(P) \to
Q$ such that the diagram
\be
\begin{array}{rcl}
 {P}  & \op\longrightarrow^{J^k} &  {\gj^s(P)}  \\
 {\Delta} & \searrow  \swarrow & {{\got f}^\Delta} \\ 
 & {Q} &  
\end{array}
\ee
is commutative.
\end{theo}

\begin{proof}
The proof is based on the following fact \cite{vinb}. Let $h\in
\hm_{\cA} (\cA\otimes P, Q)$ and 
\be
\wh a: P\ni p\to a\otimes p\in \cA\otimes P,
\ee
then 
\be
\dl_b(h\circ\wh a)(p) = h(\dl^b(a\otimes p)).
\ee
\end{proof}

 The correspondence $\Delta\mapsto{\got f}^\Delta$ defines
the isomorphism
\mar{5.50}\beq
\hm_{\cA}(\gj^s(P),Q)=\dif_s(P,Q), \label{5.50}
\eeq
which shows that 
the jet module  $\gj^s(P)$ is the
representative object of the functor $Q\to \dif_s(P,Q)$.

Let us consider the particular jet modules $\gj^s(\cA)$ of the algebra 
$\cA$, denoted simply by $\gj^s$. 
 The module $\gj^s$
can be provided with the structure of a commutative algebra with respect
to the multiplication
\be
(aJ^sb)\cdot(a'J^sb)= aa'J^s(bb').
\ee

For instance, the algebra $\gj^1$ consists of the elements $a\otimes b$ modulo
the relations
\mar{5.53}\beq
a\otimes b + b\otimes a = ab\otimes \bb + \bb\otimes ab.   \label{5.53}
\eeq
It has the left $\cA$-module structure 
\mar{+a6}\beq
c((a\ot b)\,{\rm mod}\,\m^2)=(ca)\ot b\,{\rm mod}\,\m^2 \label{+a6}
\eeq
(\ref{+a21}) and the $\star$-left $\cA$-module structure
\mar{+a4}\beq
c\star ((a\ot b)\,{\rm mod}\,\m^2)= a\ot (cb)\,{\rm mod}\,\m^2 \label{+a4}
\eeq
(\ref{mos057}) which coincides with the right $\cA$-module structure
(\ref{+a31}).
We have the canonical monomorphism of left $\cA$-modules
\mar{5.45}\beq
i_1: \cA \to \gj^1, \qquad i_1:a \mapsto a\otimes \bb\, {\rm
mod}\,\m^2, \label{5.45}
\eeq
and the corresponding projection 
\mar{+216}\ben
&&\gj^1\to\gj^1/ \im i_1=(\Ker \m^1){\rm mod}\,\m^2=\cO^1, \label{+216}\\
&& a\ot b\,{\rm mod}\,\m^2\to (a\ot b-ab\ot \bb)\,{\rm mod}\,\m^2. \nonumber
\een
The quotient $\cO^1$ (\ref{+216}) consists of the elements
\be
(a\ot b-ab\ot \bb)\,{\rm mod}\,\m^2,
\qquad \forall a,b\in\cA.
\ee
It is provided both with the central $\cA$-bimodule structure
\mar{+a1,2}\ben
&& c((a\ot b-ab\ot \bb)\,{\rm mod}\,\m^2)=(ca\ot b-cab\ot \bb)\,{\rm
mod}\,\m^2, \label{+a1}\\
&& ((\bb\ot ab - b\ot a) \,{\rm mod}\,\m^2)c=(\bb\ot abc -b\ot ac) \,{\rm
mod}\,\m^2 \label{+a2}
\een
and  the $\star$-left $\cA$-module structure 
\mar{+a3}\beq
c\star ((a\ot b-ab\ot \bb)\,{\rm mod}\,\m^2)=(a\ot cb-acb\ot \bb)\,{\rm
mod}\,\m^2. \label{+a3}
\eeq
It is readily observed that the projection (\ref{+216}) is both the left and
$\star$-left module morphisms.
Then we have the $\star$-left module morphism
\mar{mos045}\ben
&&d^1: \cA \op\to^{J^1} \gj^1\to\cO^1, \label{mos045}\\
&& d^1: b\to \bb\ot b\,{\rm mod}\,\m^2\to (\bb\ot b- b\ot \bb)\,
{\rm mod}\,\m^2,
\nonumber
\een
such that the central $\cA$-bimodule $\cO^1$ is generated by the elements
$d^1(b)$, $b\in\cA$, in accordance with the law
\mar{+896}\beq
ad^1b=(a\ot b -ab\ot\bb) \,{\rm mod}\,\m^2=(\bb\ot ab) -b\ot a)\,{\rm
mod}\,\m^2=(d^1b)a. \label{+896}
\eeq

\begin{prop} \label{ch527} \mar{ch527}
The morphism $d^1$ (\ref{mos045}) is a derivation from $\cA$ to $\cO^1$ seen
both as a left $\cA$-module and $\cA$-bimodule.
\end{prop}

\begin{proof} 
Using the relations (\ref{5.53}), one obtains in an explicit
form that 
\mar{+850}\ben
&& d^1(ba)= (\bb\otimes ba - ba\otimes \bb)\,{\rm mod}\,\m^2 =
\nonumber\\ 
&& \qquad (b\otimes a  +a\otimes b  - ba\otimes
\bb-ab\otimes
\bb)\,{\rm mod}\,\m^2  =bd^1a + ad^1b. \label{+850}
\een
This is a $\cO^1$-valued first order differential operator. At the same time,
\be
d^1(ba)=(\bb\otimes ba - ba\otimes \bb +b\ot a - b\ot a)\,{\rm mod}\,\m^2 =
  (d^1b)a + bd^1a.
\ee
\end{proof}

With the derivation $d^1$ (\ref{mos045}), we get the left and $\star$-left
module splitting 
\mar{mos058,070}\ben
&&\gj^1=\cA\oplus \cO^1, \label{mos058} \\
&& a\gj^1(cb)= ai_1(cb) +a d^1(cb). \label{mos070}
\een
Accordingly, there is the exact sequence 
\mar{+440} \beq
0\to \cO^1 \to \gj^1 \to \cA \to 0 \label{+440}
\eeq
which is split by the monomorphism (\ref{5.45}).

\begin{prop} \label{mos073} \mar{mos073} There is the isomorphism
\mar{mos074}\beq
\gj^1(P)=\gj^1\ot P, \label{mos074}
\eeq
where by $\gj^1\ot P$ is meant the tensor product of the right ($\star$-left)
$\cA$-module
$\gj^1$ (\ref{+a4}) and the left $\cA$-module $P$, i.e.,
\be
[a\ot b\,{\rm mod}\, \m^2]\ot p=[a\ot\bb \,{\rm mod}\, \m^2]\ot bp.
\ee
\end{prop}

\begin{proof}
The isomorphism (\ref{mos074}) is 
given by the assignment
\mar{+170}\beq
(a\ot bp)\,{\rm mod}\,\m^2 \leftrightarrow 
[a\ot b\,{\rm mod}\,\m^2]\ot p. \label{+170}
\eeq
\end{proof}

The isomorphism (\ref{mos058}) leads to 
the isomorphism
\be
&& \gj^1(P)= (\cA\oplus \cO^1)\ot P, \\
&& (a\ot bp)\,{\rm mod}\,\m^2 \leftrightarrow  [(ab +ad^1(b))\,{\rm
mod}\,\m^2]\ot p, 
\ee
and to the splitting of left and $\star$-left $\cA$-modules
\mar{mos071}\beq
\gj^1(P)= (\cA \ot P)\oplus (\cO^1\ot P), \label{mos071}
\eeq
Applying the projection $\pi^1_0$ (\ref{+a13}) to the splitting
(\ref{mos071}), we obtain the exact sequence of left and $\star$-left
$\cA$-modules (\ref{+175})
\be
 && 0\to [(a\ot b -ab\ot \bb)\,{\rm mod}\,\m^2]\ot p\to
[(c\ot \bb+ a\ot b -ab\ot \bb)\,{\rm mod}\,\m^2]\ot p  \\
&& \qquad = (c\ot p +a\ot bp
-ab\ot p)\,{\rm mod}\,\m^2 \to cp,
\ee
similar to the exact sequence (\ref{+440}). This exact sequence has
the canonical splitting by the $\star$-left $\cA$-module morphism
\be
P\ni ap \mapsto a\ot p + d^1(a)\ot p.
\ee
However,  the exact
sequence (\ref{+175}) needs not be split by a left
$\cA$-module morphism. Its splitting by a left $\cA$-module morphism (see
(\ref{+177}) below) implies a connection. On can treat the canonical
splitting (\ref{5.45}) of the exact sequence (\ref{+440}) as being the
canonical connection on the algebra $\cA$.

In the case of  $\gj^s$, the isomorphism (\ref{5.50}) takes the form
\mar{5.52}\beq
\hm_{\cA}(\gj^s,Q)=\dif_s(\cA,Q). \label{5.52}
\eeq
Then Theorem \ref{ch526} and Proposition \ref{ch527} lead to the isomorphism 
\mar{5.56}\beq
\hm_{\cA}(\cO^1,Q)=\gd(\cA,Q). \label{5.56}
\eeq
In other words, any $Q$-valued derivation of $\cA$ is represented by the
composition  $h\circ d^1$, $h\in \hm_{\cA}(\cO^1,Q)$, due to the property 
$d^1(\bb)= 0$.

For instance,
if $Q=\cA$, the isomorphism (\ref{5.56}) reduces to the duality relation
\mar{5.81}\ben
&& \hm_{\cA}(\cO^1,\cA)=\gd(\cA), \label{5.81}\\
&&  u(a)=u(d^1a), \qquad a\in\cA, \nonumber
\een
i.e., the module $\gd\cA$ coincides with the left $\cA$-dual $\cO^{1*}$
of $\cO^1$.

Let us define the modules $\cO^k$ as the skew tensor products
of the $\cK$-modules  $\cO^1$.

\begin{prop}\label{ch528} \mar{ch528} \cite{vinb}.
 There are the isomorphisms 
\mar{5.57,'}\ben
&& \hm_{\cA}(\cO^k,Q)=\gd_k(\cA,Q), \label{5.57}\\
&& \hm_\cA(\gj^1(\cO^k), Q)= \gd_k(\dif_1(Q)). \label{5.57'}  
\een
\end{prop} 

The isomorphism (\ref{5.57}) is the higher order extension of the isomorphism
(\ref{5.56}). It shows that the module
$\cO^k$ is a representative object of the derivation functor $Q\to
\gd_k(\cA,Q)$. 

The isomorphism (\ref{5.57'}) implies  the homomorphism
\be
h^k: \gj^1(\cO^{k-1})\to \cO^k
\ee
and defines the operators of exterior differentiation
\mar{5.4}\beq
d^k=h^k\circ J^1: \cO^{k-1}\to \cO^k. \label{5.4}
\eeq
These operators constitute the De 
Rham complex
\mar{55.63}\beq
0\longrightarrow \cA\op\longrightarrow^{d^1}\cO^1\op\longrightarrow^{d^2}
\cdots  \cO^k\op\longrightarrow^{d^{k+1}} \cdots .
\label{55.63}
\eeq

\section{Connections on commutative modules}

There are several equivalent definition of connections on modules over a
commutative ring.

\begin{defi} \label{+176} \mar{+176}
By a  connection on a $\cA$-module $P$ is
called a left
$\cA$-module morphism 
\mar{+177,9'}\ben
&&\G:P\to \gj^1(P), \label{+177}\\
&& \G (ap)=a\G(p), \label{+179'}
\een
which splits the exact sequence (\ref{+175}).
\end{defi}

This splitting reads 
\mar{+178}\beq
J^1p=\G(p) + \nabla^\G(p), \label{+178}
\eeq
where $\nabla^\G$ is the complementary morphism
\mar{+179}\ben
&& \nabla^\G: P\to \cO^1\ot P, \label{+179} \\
&& \nabla^\G(p)= \bb\ot p\,{\rm mod}\, \m^2- \G(p). \nonumber
\een
This complementary morphism makes the sense of a  covariant
differential on the module $P$,
but we will follow the tradition to use the terms "covariant differential"
and "connection" on modules synonymously.
With the relation (\ref{+179'}), we find
that $\nabla^\G$ obeys the  Leibniz rule
\mar{+180}\beq
\nabla^\G(ap)= da \ot p +a\nabla^\G(p). \label{+180}
\eeq

\begin{defi} \label{+181} \mar{+181}
By a  connection on a $\cA$-module $P$ is meant any morphism $\nabla$
(\ref{+179}) which obeys the Leibniz rule (\ref{+180}), i.e., $\nabla$ is  a
$(\cO^1\ot P)$-valued first order differential operator on $P$. 
\end{defi}

In view of Definition (\ref{+181}) and of the isomorphism (\ref{mos071}), it is
more convenient to rewrite the exact sequence (\ref{+175}) into the form
\mar{+183}\beq
0\to \cO^1\ot P\to (\cA\oplus \cO^1)\ot P\to P\to 0. \label{+183}
\eeq
Then a connection $\nabla$ on $P$ can be defined as a left
$\cA$-module splitting of this exact sequence.

In the case of the ring $C^\infty(X)$ and a locally free $C^\infty(X)$-module
$\cS$ of finite rank, there exist  the isomorphisms 
\mar{+844}\ben
&& \cO^1(X)=\hm_{C^\infty(X)}(\gd(C^\infty(X)),C^\infty(X)), \label{+844}\\
&&  \hm_{C^\infty(X)}(\gd(C^\infty(X)),\cS)= \cO^1(X)\ot \cS. \nonumber
\een
With these isomorphisms, 
we come to other equivalent definitions of a connection on modules.

\begin{defi} \label{mos082} \mar{mos082}
Any morphism
\mar{mos083}\beq
\nabla: \cS \to \hm_{C^\infty(X)}(\gd(C^\infty(X)),\cS) 
\eeq
 satisfying the Leibniz rule (\ref{+180}) is called a connection
 on a
$C^\infty(X)$-module $\cS$.
\end{defi}

\begin{defi} \label{mos086} \mar{mos086}
By a connection on a
$C^\infty(X)$-module $\cS$ is meant 
a $C^\infty(X)$-module morphism
\mar{mos089}\beq
\gd(C^\infty(X))\ni \tau\mapsto \nabla_\tau\in \dif_1(\cS,\cS)
\label{mos089}
\eeq
such that the first order differential operators $\nabla_\tau$ obey the 
rule
\mar{mos087}\beq
\nabla_\tau (fs)= (\tau\rfloor df)s+ f\nabla_\tau s. \label{mos087}
\eeq
\end{defi}

If a $\cS$ is a commutative $C^\infty(X)$-ring, Definition \ref{mos086} can be
modified as follows.

\begin{defi} \label{mos088} \mar{mos088}
By a  connection on $C^\infty(X)$-ring $\cS$
is meant any
$C^\infty(X)$-module morphism
\mar{mos090}\beq
\gd(C^\infty(X))\ni \tau\mapsto \nabla_\tau\in \gd\cS \label{mos090}
\eeq
which is a connection on $\cS$ as a $C^\infty(X)$-module, i.e., obeys the
Leinbniz rule (\ref{mos087}).
\end{defi}

Two such
connections $\nabla_\tau$ and $\nabla'_\tau$ differ from each other in a
derivation of the ring $\cS$ which vanishes on $C^\infty(X)\subset \cS$.

\section{Noncommutative differential calculus}

One believes that a noncommutative generalization of differential
geometry should be given by a $\Bbb Z$-graded differential algebra
which replaces 
the exterior algebra of differential forms \cite{malt}. This viewpoint is more
general than that implicit above where a noncommutative ring
replaces a ring of smooth functions. 

Recall that a  graded algebra
$\Om^*$ over a commutative ring $\cK$ is defined as a direct sum  
\be
\Om^*= \op\oplus_{k=0} \Om^k 
\ee
of $\cK$-modules $\Om^k$, provided with the associative multiplication
law such that
$\al\cdot\bt\in \Om^{|\al|+|\bt|}$, where $|\al|$ denotes the degree of
an element 
$\al\in \Om^{|\al|}$. In particular, $\Om^0$ is a unital $\cK$-algebra
$\cA$, while 
$\Om^{k>0}$ are $\cA$-bimodules. 
A graded algebra $\Om^*$ is called a 
graded differential algebra if it is
a cochain complex of $\cK$-modules
\be
0\ar \cA\ar^\dl\Om^1\ar^\dl\cdots
\ee
with respect to a coboundary operator $\dl$ such that
\be
\dl(\al\cdot\bt)=\dl\al\cdot\bt +(-1)^{|\al|}\al\cdot \dl\bt.
\ee

A graded differential algebra $(\Om^*,\dl)$ with $\Om^0=\cA$ is called the
 differential calculus over $\cA$.
If $\cA$ is a $*$-algebra, we have additional conditions
\be
&& (\al\cdot\bt)^*=(-1)^{|\al||\bt|}\bt^*\al^*,\\
&& (\dl\al)^*=\dl(\al^*).
\ee

\begin{rem} \label{+935} \mar{+935} 
The De Rham complex (\ref{55.63}) exemplifies a differential calculus over a
commutative ring.  
To generalize it to a noncommutative ring $\cA$, 
the coboundary operator $\dl$ should have the additional properties:
\begin{itemize}
\item $\Om^{k>0}$ are central $\cA$-bimodules,
\item elements $\dl a_1\cdots \dl a_k$, $a_i\in\cZ(\cA)$, belong to the
centre $\cZ(\Om^k)$ of the module $\Om^k$. Then, if $\cA$ is a commutative
ring, the commutativity condition (\ref{+896}) holds.
\end{itemize}
\end{rem}

Let $\Om^*\cA$ be the smallest differential subalgebra of the
algebra $\Om^*$ which contains $\cA$. As an $\cA$-algebra, it is
generated by the elements  
$\dl a$, $a\in \cA$, and consists of finite linear combinations of
monomials of the form
\mar{+892}\beq
\al=a_0\dl a_1\cdots \dl a_k, \qquad a_i\in \cA. \label{+892}
\eeq
The product of monomials (\ref{+892}) is defined by 
the rule 
\be
(a_0\dl a_1)\cdot (b_0\dl b_1)=a_0\dl (a_1b_0)\cdot \dl b_1- a_0a_1\dl
b_0\cdot \dl b_1. 
\ee
In particular, $\Om^1\cA$ is a
$\cA$-bimodule generated by elements $\dl a$, $a\in A$. 
Because of
\be
(\dl a)b=\dl (ab)-a\dl b,
\ee
the bimodule $\Om^1\cA$ can also be seen as a left [right]
$\cA$-module generated by the elements $\dl a$,
$a\in\cA$. Note that $\dl(\bb)=0$.
Accordingly, 
\be
\Om^k\cA=\underbrace{\Om^1\cA\cdots \Om^1\cA}_k
\ee
are $\cA$-bimodules and, simultaneously, left [right] $\cA$-modules
generated by monomials (\ref{+892}).

The differential subalgebra $(\Om^*\cA,\dl)$ is a differential
calculus over $\cA$. It is called the  universal differential
calculus because of the
following property \cite{conn0,karo1,land}. 
Let $(\Om'^*,\dl')$ be another
differential calculus over a unital $\cK$-algebra $\cA'$, and let
$\rho:\cA\to \cA'$ be an algebra morphism. There exists a unique
extension of this morphism to a morphism of graded differential algebras
\be
\rho^k: \Om^k\cA \to \Om'^k
\ee
such that $\rho^{k+1}\circ \dl=\dl' \circ\rho^k$.

Our interest to differential calculi over an algebra $\cA$ is
caused by the fact that, in commutative geometry, Definition \ref{+181} of a
connection 
on an $\cA$-module requires the module $\cO^1$ (\ref{+216}). If
$\cA=C^\infty(X)$, this module is
the module of 1-forms on $X$. 
To introduce connections in
noncommutative geometry, one therefore should construct the
noncommutative version of the module $\cO^1$. 
We may follow the construction of $\cO^1$ in Section 3, but not take
the quotient by mod$\m^2$ that implies  the
commutativity condition (\ref{+896}). 

\begin{rem}
This is the crucial poin that does not
enable us to generalize the notion of jets of modules to modules over a
noncommutative ring unless the very particular case when $d\cA$ belongs to
the centre of the module $\Om^1$.
\end{rem}

Given a unital
$\cK$-algebra
$\cA$, let us consider the tensor product
$\cA\op\ot_\cK\cA$ of $\cK$-modules and the $\cK$-module morphism 
\be
\m^1:\cA\op\ot_\cK\cA\ni a\ot b\mapsto ab\in\cA.
\ee 
Following (\ref{+216}), we define the $\cK$-module
\mar{+891}\beq
\ol\cO^1[\cA]=\Ker \m^1. \label{+891}
\eeq
There is
the $\cK$-module morphism 
\mar{+898}\beq
d:\cA\ni a\mapsto (\bb\ot a - a\ot \bb)\in \ol\cO^1[\cA]
\label{+898}
\eeq
(cf. (\ref{mos045})).  
Moreover, $\ol\cO^1[\cA]$
is a $\cA$-bimodule generated by the elements $da$, $a\in A$, with the
multiplication law
\be
b(da)c=b\ot ac- ba\ot c, \qquad a,b,c\in\cA.
\ee
The morphism $d$ (\ref{+898}) possesses the property
\mar{+851}\beq
d(ab)= (\bb\ot ab- ab\ot \bb + a\ot b -a\ot b)= (da)b +adb
\label{+851} 
\eeq
(cf. (\ref{+850})), i.e., $d$ is a $\ol\cO^1[\cA]$-valued derivation of $\cA$.
Due to this property, $\ol\cO^1[\cA]$ can be seen as a
left $\cA$-module generated by the elements  $da$, $a\in\cA$. At the same
time, if $\cA$ is a commutative ring, the $\cA$-bimodule $\ol\cO^1[\cA]$ does
not coincide with the bimodule $\cO^1$ (\ref{+216}) because $\ol\cO^1[\cA]$
is not a central bimodule (see Remark \ref{+935}).

To overcome this difficulty, let us consider the $\cZ(\cA)$ of derivations of
the algebra $\cA$. They obey the rule 
\mar{+831}\beq
u(ab)=u(a)b + au(b), \qquad \forall a,b\in \cA. \label{+831}
\eeq
It should be emphasized that the derivation rule (\ref{+831}) differs
from that
\be
u(ab)=u(a)b +u(b)a
\ee
for a general algebra \cite{lang}. By virtue of
(\ref{+831}), derivations of an algebra $\cA$ constitute a $\cZ(\cA)$-bimodule,
but not a left $\cA$-module.

The $\cZ(\cA)$-bimodule $\gd\cA$ is also a Lie algebra over the
commutative ring $\cK$ with respect to
the Lie bracket 
\mar{+860}\beq
[u,u']=u\circ u' - u'\circ u. \label{+860}
\eeq
The centre $\cZ(\cA)$ is stable under $\gd\cA$, i.e.,
\be
u(a)b=bu(a), \quad \forall a\in\cZ(\cA),\quad b\in\cA, \quad u\in\gd\cA,
\ee
and one has
\mar{+834}\beq
[u,au']=u(a)u' +a[u,u'], \quad \forall a\in \cZ(\cA), \quad
u,u'\in\gd\cA. \label{+834}
\eeq
If $\cA$ is a unital $*$-algebra, the module $\gd\cA$ of derivations
of $\cA$ is provided with the involution $u\mapsto u^*$ defined by
\be
u^*(a)=(u(a^*))^*.
\ee
Then the Lie bracket (\ref{+860}) satisfies the reality condition
$[u,u']^*=[u^*,u'^*]$. 

Let us consider the  Chevalley--Eilenberg cohomology (see
\cite{vais})  of the Lie algebra $\gd\cA$
with respect to its natural representation in $\cA$. The corresponding
$k$-cochain space $\ul\cO^k[\cA]$, $k=1,\ldots,$ is the
$\cA$-bimodule of $\cZ(\cA)$-multilinear antisymmetric mappings of
$\gd\cA^k$ to $\cA$. In particular, $\ul\cO^1[\cA]$ is the $\cA$-dual
\mar{+861}\beq
\ul\cO^1[\cA]=\gd\cA^* \label{+861}
\eeq
of the derivation module $\gd\cA$ (cf. (\ref{+844})).
Put $\ul\cO^0[\cA]=\cA$. The  Chevalley--Eilenberg
coboundary operator 
\be
d: \ul\cO^k[\cA]\to \ul\cO^{k+1}[\cA]
\ee
is given by
\mar{+840}\ben
&& (d\f)(u_0,\ldots,u_k)=\frac{1}{k+1}\op\sum^k_{i=0}(-1)^iu_i
(\f(u_0,\ldots,\wh{u_i},\ldots,u_k)) +\label{+840}\\
&& \qquad \frac{1}{k+1}\op\sum_{0\leq r<s\leq k} (-1)^{r+s}
\f([u_r,u_s],u_0,\ldots,
\wh{u_r}, \ldots, \wh{u_s},\ldots,u_k), \nonumber
\een
where $\wh{u_i}$ means omission of $u_i$. For instance,
\mar{+920,1}\ben
&& (da)(u)=u(a), \qquad a\in\cA, \label{+920}\\
&& (d\f)(u_0,u_1)= \frac12(u_0(\f(u_1)) -u_1(\f(u_0)) -\f([u_0,u_1])), \quad
\f\in \cO^1[\cA]. \label{+921}
\een
It is readily observed that
$d^2=0$, and we have the  Chevalley--Eilenberg 
cochain complex of $\cK$-modules
\mar{+841}\beq
0\ar \cA\ar^d \ul\cO^k[\cA] \ar^d \cdots. \label{+841}
\eeq
Furthermore, the $\Bbb Z$-graded space
\mar{+842}\beq
\ul\cO^*[\cA]=\op\oplus_{k=0} \ul\cO^k[\cA] \label{+842}
\eeq
is provided with the structure of a graded algebra with respect to
the multiplication $\w$ combining the product of $\cA$ with
antisymmetrization in 
the arguments. Notice that, if $\cA$ is not commutative, there is
nothing like graded commutativity of forms, i.e.,
\be
\f\w\f'\neq (-1)^{|\f||\f'|}\f'\w\f
\ee
in general. If $\cA$ is a $*$-algebra, $\ul\cO^*[\cA]$ is also
equipped with the involution 
\be
\f^*(u_1,\ldots,u_k)\op=^{\rm def}(\f(u^*_1,\ldots,u^*_k))^*.
\ee
Thus, $(\ul\cO^*[\cA],d)$ is a differential calculus over $\cA$,
called the  Chevalley--Eilenberg differential calculus.

It is easy to see that, if $\cA=\Bbb C^\infty(X)$ is the
commutative ring of smooth complex functions on a compact manifold $X$, the
graded algebra $\ul\cO^*[\Bbb C^\infty(X)]$ is exactly the complexified
exterior algebra $\Bbb C\ot\cO^*(X)$ of exterior forms on $X$. 
In this case, the coboundary operator (\ref{+840}) coincides with the exterior
differential, and (\ref{+841}) is the De Rham complex of complex
exterior forms on a manifold $X$.
In particular,
the operations
\be
&& (u\rfloor\f)(u_1,\ldots,u_{k-1})= k\f(u,u_1,\ldots,u_{k-1}), \qquad
u\in\gd\cA, \\
&& \bL_u(\f)=d(u\rfloor\f) +u\rfloor f(\f),
\ee 
are the noncommutative generalizations of the contraction and the Lie
derivative of differential forms. 
These
facts motivate one to think of elements of $\cO^1[\cA]$ as being a
noncommutative generalization of differential 1-forms, though
this generalization by no means is unique.

Let $\cO^*[\cA]$ be the smallest differential subalgebra of the
algebra $\ul\cO^*[\cA]$ which contains $\cA$. It is generated by the elements 
$da$, $a\in \cA$, and consists of finite linear combinations of
monomials of the form
\be
\f=a_0da_1\w\cdots \w da_k, \qquad a_i\in \cA,
\ee
(cf. (\ref{+892})).
In particular, $\cO^1[\cA]$ is a 
$\cA$-bimodule (\ref{+891}) generated by $da$, $a\in\cA$. Since the
centre $\cZ(\cA)$ of
$\cA$ is
stable under derivations of $\cA$, we have
\be
&& bda=(da)b, \qquad adb=(db)a, \qquad a\in \cA, \qquad b\in \cZ(\cA), \\
&& da\w db=-db\w da, \qquad \forall a\in \cZ(\cA).
\ee
Hence, $\cO^1[\cA]$ is a central bimodule in contrast with the bimodule
$\ol\cO^1[\cA]$ (\ref{+891}). 
 By virtue of the relation (\ref{+920}),  
we have the isomorphism 
\mar{+853}\beq
\gd\cA=\cO^1[\cA]^* \label{+853}
\eeq
of the $\cZ(\cA)$-module $\gd\cA$ of derivations
of $\cA$ to the $\cA$-dual of the module $\cO^1[\cA]$ (cf. (\ref{5.81})).
Combining the duality relations (\ref{+861}) and (\ref{+853}) gives
the relation
\be
\ul\cO^1[\cA]=\cO^1[\cA]^{**}.
\ee

The differential subalgebra $(\cO^*[\cA],d)$ is a universal differential
calculus over $\cA$. If $\cA$ is a commutative ring, then
$\cO^*[\cA]$ is the De Rham complex (\ref{55.63}).

\section{Universal connections}

Let $(\Om^*,\dl)$ be a differential calculus over a unital $\cK$-algebra
$\cA$ and $P$ a left [right] $\cA$-module.
Similarly to Definition
\ref{+181}, one can construct the
tensor product $\Om^1\ot P$ [$P\ot\Om^1$] and define a connection on
$P$ as follows \cite{var,land}.  

\begin{defi} \label{+864} \mar{+864}
A  noncommutative connection on a left
$\cA$-bimodule $P$  with respect to the differential calculus $(\Om^*,\dl)$
is a $\cK$-module morphism 
\mar{+866}\beq
\nabla: P\to \Om^1\ot P \label{+866}
\eeq
which obeys the Leibniz rule
\be
\nabla(ap)=\dl a\ot p +a\nabla(p).
\ee
\end{defi}

If $\Om^*=\Om^*\cA$ is a universal differential calculus, the
connection (\ref{+866}) is called a  universal connection \cite{var,land}. 

The  curvature
of the noncommutative connection (\ref{+866}) is  
defined as the $\cA$-module morphism
\be
\nabla^2: P\to \cO^2[\cA]\ot P 
\ee
 \cite{land}.
Note also that the morphism (\ref{+866}) has a natural extension
\be
&& \nabla: \Om^k\ot P\to \Om^{k+1}\ot P, \\
&& \nabla(\al\ot p)=\dl\al\ot p +(-1)^{|\al|}\al\ot\nabla(p), \qquad \al\in
\Om^*, 
\ee
\cite{land,dub2}. 

Similarly, a noncommutative connection on a right $\cA$-module is defined. 
However, a connection on a left [right] module does not necessarily
exist as it is illustrated by the following theorem.

\begin{theo} \label{+893} \mar{+893}
 A left [right] universal connection on a left [right] module $P$
 of finite rank exists if and only if $P$ is projective
\cite{land,cunt}.
\end{theo}

The problem arises when $P$ is a $\cA$-bimodule. If $\cA$ is a
commutative ring, left and right module structures of an $\cA$-bimodule
are equivalent, and one deals with either a left or right
noncommutative connection on
$P$ (see Definition \ref{+181}). If $P$ is a $\cA$-bimodule over a
noncommutative ring,  left and  right connections $\nabla^L$ and
$\nabla^R$ on
$P$ should be considered simultaneously. However, the 
pair $(\nabla^L,\nabla^R)$ by no means is a bimodule connection since
$\nabla^L(P)\in \Om^1\ot P$, whereas $\nabla^R(P)\in P\ot\Om^1$. As a
palliative, one assumes that there exists a bimodule isomorphism
\mar{+894}\beq
\varrho: \Om^1\ot P\to P\ot\Om^1. \label{+894}
\eeq
Then a pair $(\nabla^L,\nabla^R)$ of right and left noncommutative
connections on $P$ 
is called a $\varrho$-compatible if 
\be
\varrho\circ\nabla^L= \nabla^R
\ee
\cite{land,dub2,mour} (see also \cite{dabr96} for a weaker condition).
Nevertheless, this is not a true bimodule connection (see the condition
(\ref{+895}) below).

\begin{rem}
If $\cA$ is a commutative ring, the isomorphism $\varrho$ (\ref{+890})
is naturally the permutation 
\be
\varrho: \al\ot p\mapsto p\ot\al, \qquad \forall \al\in\Om^1, \quad p\in
P.
\ee
\end{rem}

The above mentioned problem of a bimodule connection is not simplified
radically even if $P=\Om^1$, together with the natural permutations
\be
\f\ot\f'\mapsto \f'\ot\f, \qquad \f,\f'\in \Om^1,
\ee
\cite{dub,mour}.

Let now $(\cO^*[\cA],d)$ be the universal differential calculus over a
noncommutative $\cK$-ring $\cA$. 
Let 
\mar{+910}\ben
&& \nabla^L : P\to \cO^1[\cA]\ot P, \label{+910}\\
&&\nabla^L(ap)=da\ot p +a\nabla^L(p). \nonumber
\een
be a left universal connection on a left $\cA$-module $P$ (cf.
Definition \ref{+181}). Due to the duality relation (\ref{+853}),
there is the $\cK$-module endomorphism
\mar{+911}\beq
\nabla_u^L: P\ni p\to u\rfloor \nabla^L(p)\in P \label{+911}
\eeq
of $P$ for any derivation $u\in \gd\cA$. If $\nabla^R$ is a right universal
connection on a right $\cA$-module $P$,  the similar endomorphism
\mar{+912}\beq
\nabla_u^R: P\ni p\to \nabla^L(p)\lfloor u \in P \label{+912}
\eeq
takes place for any derivation $u\in \gd\cA$. Let
$(\nabla^L,\nabla^R)$ be a $\varrho$-compatible pair of left and right
universal connections on an $\cA$-bimodule $P$. It seems natural to say
that this pair is a bimodule universal connection on $P$ if
\mar{+895}\beq
u\rfloor\nabla^L(p)= \nabla^R(p)\lfloor u \label{+895}
\eeq
for all $p\in P$ and $u\in \gd\cA$. Nevertheless, motivated by the
endomorphisms (\ref{+911}) -- (\ref{+912}), one can suggest another
definition of connections on a bimodule, similar to Definition \ref{mos086}.

\section{The Dubois-Violette connection}

Let $\cA$ be $\cK$-ring and $P$ an $A$-module of
type $(i,j)$ in accordance with the notation in Section 2.

\begin{defi} \label{+845} \mar{+845}
By analogy with Definition
\ref{mos086}, a  Dubois-Violette connection on 
an $A$-module $P$ of type $(i,j)$ is a $\cZ(\cA)$-bimodule morphism  
\mar{+846}\beq
\nabla: \gd\cA\ni u\mapsto \nabla_u\in \hm_\cK (P,P) \label{+846}
\eeq
of $\gd\cA$ to the $\cZ(\cA)$-bimodule of endomorphisms of the
$\cK$-module $P$ which obey the Leibniz rule
\mar{+847}\beq
\nabla_u(a_ipa_j)=u(a_i)pa_j +a_i\nabla_u(p)a_j +a_ipu(a_j), \quad
\forall p\in P, \quad 
\forall a_k\in A_k, \label{+847}
\eeq
\cite{dub,mour}.
\end{defi}

By virtue
of the duality relation (\ref{+853}) and the expressions (\ref{+911})
-- (\ref{+912}), every left [right] universal connection yields a
connection (\ref{+846}) on a left [right] $\cA$-module $P$. From now
on, by a connection in noncommutative geometry is meant a
Dubois-Violette connection in accordance with Definition (\ref{+845}). 

A glance at the expression (\ref{+847})
shows that, if connections on an $A$-module $P$ of type $(i,j)$ exist,
they constitute an affine space 
modelled over the linear space of $\cZ(\cA)$-bimodule morphisms
\be
\si: \gd\cA\ni u\mapsto \si_u\in \hm_{A_i-A_j}(P,P)
\ee
of $\gd\cA$ to the $\cZ(\cA)$-bimodule of endomorphisms 
\be
\si_u(a_ipa_j)=a_i\si(p)a_j, \quad
\forall p\in P, \quad 
\forall a_k\in A_k,
\ee
of the $A$-module $P$. 

\begin{ex}
If $P=\cA$, the morphisms
\mar{+870}\beq
\nabla_u(a)=u(a), \quad \forall u\in \gd\cA, \quad \forall a\in \cA,
\label{+870} 
\eeq
define a canonical connection on $\cA$ in
accordance with Definition \ref{+845}. Then the Leibniz rule
(\ref{+847}) shows that any connection on a central $\cA$-bimodule $P$
is also a connection on $P$ seen as a $\cZ(\cA)$--bimodule.
\end{ex}

\begin{ex}
If $P$ is a $\cA$-bimodule and $\cA$ has only inner derivations 
\be
{\rm ad}\, b(a)= ba-ab,
\ee
the morphisms
\mar{+871}\beq
\nabla_{{\rm ad} b}(p)= bp-pb, \quad \forall b\in\cA, \quad \forall
p\in P, \label{+871}
\eeq
define a canonical connection on $P$.
\end{ex}

By the  curvature $R$ 
of a connection $\nabla$
(\ref{+846}) on an $A$-module $P$ is meant
the $\cZ(\cA)$-module morphism
\mar{+874}\ben
&& R:\gd\cA\times\gd\cA \ni(u,u')\to R_{u,u'}\in
\hm_{A_i-A_j}(P,P), \label{+874}\\ 
&& R_{u,u'}(p)=\nabla_u(\nabla_{u'}(p)) -\nabla_{u'}(\nabla_u(p))
-\nabla_{[u,u']}(p), \quad p\in P,\nonumber
\een
 \cite{dub}. We have
\be
&& R_{au,a'u'}=aa' R_{u,u'}, \qquad a,a'\in\cZ(\cA),\\
&& R_{u,u'}(a_ipb_j)= a_iR_{u,u'}(p)b_j, \quad a_i\in A_j, \quad b_j\in
A_j.
\ee
For instance, the curvature of the connections (\ref{+870}) and
(\ref{+871}) vanishes.

Let us provide some standard operations with the connections
(\ref{+846}). 

(i) Given two modules $P$ and $P'$ of the same type $(i,j)$ and
connections $\nabla$ and $\nabla'$ on them, there is an obvious
connection $\nabla\oplus\nabla'$ on $P\oplus P'$.

(ii) Let $P$ be a module of type $(i,j)$ and $P^*$ its $\cA$-dual. For
any connection $\nabla$ on $P$, there is a unique dual connection
$\nabla'$ on $P^*$ such that
\be
u(\lng p,p'\rng)=\lng \nabla_u(p),p'\rng +\lng p,\nabla'(p')\rng, \quad
 p\in P,\quad\ p'\in P^*, \quad u\in\gd\cA.
\ee

(iii) Let $P_1$ and $P_2$ be $A$-modules of types $(i,k)$ and $(k,j)$,
respectively, and let $\nabla^1$
and $\nabla^2$ be connections on these modules. For any $u\in\gd\cA$,
let us consider the endomorphism 
\mar{+848}\beq
(\nabla^1\ot\nabla^2)_u= \nabla_u^1\ot \id P_1 + \id P_2\ot\nabla_u^2
\label{+848} 
\eeq
of the tensor product $P_1\ot P_2$ of $\cK$-modules $P_1$ and $P_2$.
This endomorphism preserves the subset of $P_1\ot P_2$ generated by
elements 
\be
p_1a\ot p_2-p_1\ot ap_2,
\ee
with $p_1\in P_1$, $p_2\in P_2$ and
$a\in A_k$. Due to this fact, the endomorphisms (\ref{+848}) define a
connection on the tensor product $P_1\ot P_2$ of modules $P_1$ and $P_2$.

(iv) If $\cA$ is a unital $*$-algebra, we have only modules of type
$(1,1)$ and $(0,0)$, i.e., $*$-modules and $\cZ(\cA)$-bimodules. Let
$P$ be a module of one of these types. If $\nabla$ is a connection on
$P$, there exists a  conjugate connection $\nabla^*$ on $P$ given by the
relation
\mar{+849}\beq
\nabla_u^*(p)=(\nabla_{u^*}(p^*))^*. \label{+849}
\eeq
A connection $\nabla$ on $P$ is said to be  real if $\nabla=\nabla^*$.

Let now $P=\ul\cO^1[\cA]$. A connection on $\cA$-bimodule
$\ul\cO^1[\cA]$ is called a 
 linear connection \cite{dub,mour}.  Note that this is not the term for an
arbitrary left [right] connection on $\ul\cO^1[\cA]$ \cite{dub2}. If
$\ul\cO^1[\cA]$ is a
$*$-module, a linear 
connection on it is assumed to be real. Given a linear connection $\nabla$ on
$\ul\cO^1[\cA]$, there is a $\cA$-bimodule homomorphism, called the 
torsion of the connection $\nabla$,
\mar{+873}\ben
&& T:\ul\cO^1[\cA]\to \ul\cO^2[\cA], \nonumber\\
&& (T\f)(u,u')=(d\f)(u,u') - \nabla_u(\f)(u') +\nabla_{u'}(\f)(u),
\label{+873}
\een
for all $u,u'\in\gd\cA$, $\f\in \ul\cO^1[\cA]$. 

\section{Matrix geometry}

This Section gives a standard example of linear connections in matrix geometry
when
 $\cA=M_n$ is the algebra of complex $(n\times n)$-matrices
\cite{dub90,mad2,mad3}. 

Let  $\{\ve_r\}$, $1\leq r\leq n^2-1$, be an anti-Hermitiam basis of the Lie
algebra $su(n)$. Elements $\ve_r$ generate $M_n$ as an algebra, while
$u_r={\rm ad}\,\ve_r$ constitute a basis of the right Lie algebra $\gd M_n$
of derivations of the algebra $M_n$, together with the commutation relations
\be
[u_r,u_q]=c^s_{rq}u_s,
\ee
where $c^s_{rq}$ are structure constants of the Lie algebra $su(n)$.
Since the centre $\cZ(M_n)$ of $M_n$
consists of matrices $\la \bb$, $\gd M_n$ is a complex free module of rank
$n^2-1$. 

Let us consider the universal differential calculus $(\cO^*[M_n],d)$ over
the algebra $M_n$, where $d$ is the Chevalley--Eilenberg coboundary
operator (\ref{+840}). There is a convenient system $\{\th^r\}$ of 
generators of $\cO^1[M_n]$ seen as a left $M_n$-module. They are given
by the relations 
\be
\th^r(u_q)= \dl^r_q\bb.
\ee
Hence, $\cO^1[M_n]$ is a free left $M_n$-module of rank $n^2-1$.
It is readily observed that elements $\th^r$ belong to the centre of
the $M_n$-bimodule $\cO^1[M_n]$, i.e., 
\mar{+930}\beq
a\th^r=\th^r a, \qquad \forall a\in M_n. \label{+930}
\eeq
It also follows that 
\mar{+931}\beq
\th^r\w\th^q=-\th^q\w\th^r. \label{+931}
\eeq
The morphism $d:M_n\to \cO^1[M_n]$ is given by the
formula (\ref{+920}). It reads
\be
d\ve_r(u_q)={\rm ad}\,\ve_q(\ve_r)=c^s_{qr}\ve_s, 
\ee
that is,
\mar{+922}\beq
d\ve_r=c^s_{qr}\ve_s\th^q. \label{+922}
\eeq
The formula (\ref{+921}) leads to the Maurer--Cartan equations
\mar{+934}\beq
d\th^r=-\frac12c^r_{qs}\th^q\w\th^s. \label{+934}
\eeq

If we define $\th=\ve_r\th^r$, the equality (\ref{+922}) can be
rewritten as
\be
da=a\th-\th a, \qquad \forall a\in M_n.
\ee
It follows that the $M_n$-bimodule $\cO^1[M_n]$ is generated by the 
element $\th$. Since $\gd M_n$ is a finite free module, one can show that
the $M_n$-bimodule $\cO^1[M_n]$ is isomorphic to the $M_n$-dual
$\ul\cO^1[M_n]$ of $\gd M_n$.

Turn now to connections on the $M_n$-bimodule $\cO^1[M_n]$. Such a
connection $\nabla$ is given by the relations
\mar{+932}\ben
&& \nabla_{u=c^ru_r}=c^r\nabla_r, \nonumber \\
&& \nabla_r(\th^p)=\om^p_{rq}\th^q, \qquad \om^p_{rq}\in M_n.
\label{+932} 
\een
Bearing in mind the equalities (\ref{+930}) -- (\ref{+931}), we obtain
from the Leibniz rule (\ref{+847}) that
\be
a\nabla_r(\th^p)=\nabla_r(\th^p)a, \qquad \forall a\in M_n.
\ee
It follows that elements $\om^p_{rq}$ in the expression (\ref{+932})
are proportional $\bb\in M_n$, i.e., complex numbers. Then the relations
\mar{+933}\beq
\nabla_r(\th^p)=\om^p_{rq}\th^q, \qquad \om^p_{rq}\in \Bbb C, \label{+933}
\eeq
define a linear connection on the $M_n$-bimodule $\cO^1[M_n]$.

Let us consider two examples of linear connections. 

(i) Since all derivations
of the algebra $M_n$ are inner, we have the curvature-free connection
(\ref{+871}) given by the relations 
\be
\nabla_r(\th^p)=0.
\ee
However, this connection is not torsion-free. The
expressions (\ref{+873}) and (\ref{+934}) result in
\be
(T\th^p)(u_r,u_q)=-c^p_{rq}.
\ee

(ii) One can show
that, in matrix geometry, there is a unique torsion-free linear connection 
\be
\nabla_r(\th^p)=-c^p_{rq}\th^q.
\ee

\section{Connes' differential calculus}

Connes' differential calculus is based on the notion of a spectral triple
\cite{var,land,conn,mad}.

\begin{defi}
A  spectral triple $(\cA,\cH,D)$  is given by a
$*$-algebra $\cA\subset B(\cH)$ of bounded operators on a Hilbert space
$\cH$, together with an (unbounded) self-adjoint operator $D=D^*$ on
$\cH$ with the 
following properties:
\begin{itemize}
\item the resolvent $(D-\la)^{-1}$, $\la\neq\Bbb R$, is a compact
operator on $\cH$,
\item $[D,\cA]\in B(\cH)$.
\end{itemize}
\end{defi}

The couple $(\cA,D)$ is also called a  $K$-cycle
over $\cA$. In many cases, $\cH$ is a $\Bbb Z_2$-graded Hilbert space
equipped with a projector $\G$ such that 
\be
\G D+D\G=0, \qquad [a,\G]=0, \quad \forall a\in\cA,
\ee
i.e., $\cA$ acts on $\cH$ by even operators, while $D$ is an odd operator.

Given a spectral triple $(\cA,\cH,D)$, let $(\Om^*\cA,\dl)$ be a
universal differential calculus over the algebra $\cA$. Let us
construct a representation of the graded differential algebra
$\Om^*\cA$ by bounded operators on $\cH$ when the Chevalley--Eilenberg
derivation $\dl$ (\ref{+840})  of $\cA$ is replaced with the bracket
$[D,a]$, $a\in\cA$:
\mar{+940}\ben
&& \pi: \Om^*\cA \to B(\cH), \nonumber\\
&& \pi (a_0\dl a_1\cdots \dl a_k) \op=^{\rm def} a_0[D,a_1]\cdots
[D,a_k]. \label{+940}
\een
Since
\be
[D,a]^*=-[D,a^*],
\ee
we have $\pi(\f)^*=\pi(\f^*)$, $\f\in \Om^*\cA$. At the same time,
$\pi$ (\ref{+940}) fails to be a representation of the graded
differential algebra
$\Om^*\cA$ because $\pi(\f)=0$ does not imply that $\pi(\dl\f)=0$. 
Therefore, one should construct the corresponding quotient in order to
obtain a graded differential algebra of operators on $\cH$.

Let $J_0$ be the graded two-sided ideal of $\Om^*\cA$ where
\be
J^k_0=\{\f\in\Om^k\cA\, :\, \pi(\f)=0\}.
\ee
Then it is readily observed that $J=J_0 +\dl J_0$ is a graded
differential two-sided ideal of $\Om^*\cA$. 
By  Connes' differential calculus is meant the pair $(\Om^*_D\cA, d)$ such that
\be
&& \Om^*_D\cA =\Om^*\cA/J, \\
&& d[\f]=[\dl \f],
\ee
where $[\f]$ denotes the class of $\f\in \Om^*\cA$ in $\Om^*_D\cA$. It
is a differential calculus over $\Om_D^0\cA=\cA$. Its $k$-cochain
submodule $\Om^*_D\cA$ consists of the classes of operators
\be
\op\sum_j a_0^j[D,a_1^j]\cdots [D,a_k^j], \qquad a_i^j\in\cA,
\ee
modulo the submodule of operators
\be
\{\op\sum_j [D,b_0^j][D,b_1^j]\cdots [D,b_{k-1}^j]\, :\,
\op\sum_j b_0^j[D,b_1^j]\cdots [D,b_{k-1}^j]=0\}.
\ee

Let now $P$ be a right finite projective module over the $*$-algebra
$\cA$. We aim to study a right connection on $P$ with respect to
Connes' differential calculus $(\Om^*_D\cA,d)$. As was mentioned above
in Theorem \ref{+893}, a right finite projective module has a connection. Let
us construct this connection in an explicit form.

Given a generic right finite projective module $P$ over a
complex ring $\cA$, let
\be
&& \bp: \Bbb C^N\ot_C \cA \to P, \\
&& i_P: P\to \Bbb C^N\op\ot_C \cA,
\ee
be the corresponding projection and injection, where $\op\ot_C$ denotes
the tensor product over $\Bbb C$. 
There is the chain of morphisms
\mar{+941}\beq
P\ar^{i_p} \Bbb C^N\ot\cA \ar^{\id\ot\dl}\Bbb C^N\ot \Om^1\cA\ar^{\bp}
P\ot\Om^1\cA,  \label{+941}
\eeq
where the canonical module isomorphism
\be
\Bbb C^N\op\ot_C\Om^1\cA=(\Bbb C^N\op\ot_C\cA)\ot\Om^1\cA
\ee
is used. It is readily observed that the composition (\ref{+941})
denoted briefly as $\bp\circ\dl$ is a right universal connection on the
module $P$.

Given the universal connection $\bp\circ \dl$ on a right finite
projective module $P$ over a $*$-algebra $\cA$, let us consider the morphism
\be
P\ar^{\bp\circ\dl} P\ot \Om^1\cA\ar^{\id\ot\pi}P\ot \Om^1_D\cA.
\ee
It is readily observed that this is a right connection $\nabla_0$ on the module
$P$ with respect to Connes' differential calculus. Any other right connection
$\nabla$ on on $P$ with respect to Connes' differential calculus takes
the form  
\mar{+945}\beq
\nabla = \nabla_0 + \si=(\id\ot\pi)\circ\bp\circ\dl +\si \label{+945}
\eeq
where $\si$ is an $\cA$ module morphism
\be
\si: P\to P\ot\Om^1_D\cA.
\ee
A components $\si$ of the connection $\nabla$ (\ref{+945}) is called a 
noncommutative gauge field.


\begin{thebibliography}{ederf}

\bibitem{vinb} I.Krasil'shchik, V.Lychagin and A.Vinogradov, {\sl Geometry of
Jet Spaces and Nonlinear Partial Differential Equations} (Gordon and Breach,
Glasgow, 1985). 

\bibitem{book} G.Giachetta, L.Mangiarotti and G.Sardanashvily, {\sl New
Lagrangian and Hamiltonian Methods in Field Theory} (World Scientific,
Singapore, 1997).

\bibitem{mcl} S.Mac Lane, {\sl Homology} (Springer-Verlag, Berlin, 1967).

\bibitem{dub} M.Dubois-Violette and P.Michor, Connections on central
bimodules in noncommutative differential geometry, {\sl J. Geom.
Phys.} {\bf 20} (1996) 218.

\bibitem{dixm} J.Dixmier, {\sl $C^*$-Algebras} (North-Holland, Amsterdam,
1977).

\bibitem{mich} E.Michael, {\sl Locally Multiplicatively Convex
Topological Algebras} (Am. Math. Soc., Providence, 1974).

\bibitem{swan} R.Swan, Vector bundles and projective modules, {\sl
Trans. Am. Math. Soc.} {\bf 105} (1962) 264.

\bibitem{var} J.V\'arilly and J.Grasia-Bondia, Connes' noncommutative
differential geometry and the Standard Model, {\sl J. Geom. Phys.} {\bf
12} (1993) 223.

\bibitem{kosz60} J.Koszul, {\sl Lectures on Fibre Bundles and Differential
Geometry} (Tata University, Bombay, 1960). 

\bibitem{malt} G. Maltsiniotis, Le langage des espaces et des groupes
quantiques, {\sl Commun. Math. Phys.} {\bf 151} (1993) 275.

\bibitem{conn0} A.Connes,  Non-commutative differential geometry, {\sl
Publ. I.H.E.S} {\bf 62} (1986) 257.

\bibitem{karo1} M.Karoubi, Connexion, courbures et classes
caracteristique en $K$-theorie algebrique, {\sl Can. Nath. Soc. Conf.
Proc.} {\bf 2} (1982) 19.

\bibitem{land} G.Landi, {\sl An Introduction to Noncommutative Spaces
and their Geometries}, Lect. Notes in Physics, New series m: Monographs,
{\bf 51} (Springer-Verlag, Berlin, 1997).

\bibitem{lang} S.Lang, {\sl Algebra} (Addison--Wisley, N.Y., 1993).

\bibitem{vais} I.Vaisman, {\sl Lectures on the Geometry of Poisson Manifolds}
(Birkh\"auser Verlag, Basel, 1994).

\bibitem{dub2} M.Dubois-Violette, J.Madore, T.Masson and J.Morad, On
curvature in noncommutative geometry, {\sl J. Math. Phys.} {\bf 37}
(1996) 4089.

\bibitem{cunt} J.Cuntz and D.Quillen, Algebra extension and
nonsingularity, {\sl J. Amer. Math. Soc.} {\bf 8} (1995) 251.

\bibitem{mour} J.Mourad, Linear connections in noncommutative geometry,
{\sl Class. Quant. Grav.} {\bf 12} (1995) 965.

\bibitem{dabr96} L.Dabrowski, P.Hajac, G.Lanfi and P.Siniscalco,
Metrics and pairs of left and right connections on bimodules, {\sl J.
Math. Phys.} {\bf 37} (1996) 4635.

\bibitem{dub90} M.Dubois-Violette, R.Kerner and J.Madore,
Noncommutative differential geometry of matrix algebras, {\sl J. Math.
Phys.} {\bf 31} (1990) 316.

\bibitem{mad2} J.Madore, T.Masson and J.Mourad, Linear connections on
matrix geometries, {\sl Class. Quant. Grav.} {\bf 12} (1995) 1429.

\bibitem{mad3} J.Madore, Linear connections on fuzzy manifolds, {\sl
Class. Quant. Grav.} {\bf 13} (1996) 2109.

\bibitem{conn} A.Connes, {\sl Noncommutative Geometry} (Academic Press,
N.Y., 1994).

\bibitem{mad} J.Madore, {\sl An Introduction to Noncommutative
Differential Geometry and its Physical Applications} (Cambridge Univ.
Press, Cambridge, 1995).


\end{thebibliography}
\end{document}